\documentclass[aps,prl,twocolumn,floatfix]{revtex4}
\usepackage{eurosym}
\usepackage{amsfonts}
\usepackage{mathrsfs}
\usepackage{color}
\usepackage{xspace}
\usepackage{subfigure}
\usepackage{longtable}
\usepackage{multirow}
\usepackage{tabularx}
\usepackage{amsmath}
\usepackage{amssymb}
\usepackage{graphicx}
\usepackage{dcolumn}
\usepackage{bm}
\usepackage[colorlinks,urlcolor=blue,citecolor=blue,linkcolor=blue]{hyperref}
\usepackage[normalem]{ulem}

\begin{document}

\title{Experimental realization of a non-magnetic one-way spin switch}
\author{M.~E.~Mossman$^{1}$}
\author{Junpeng Hou$^{2}$}
\author{Xi-Wang Luo$^{2}$}
\author{Chuanwei Zhang$^{2}$}
\thanks{chuanwei.zhang@utdallas.edu}
\author{P.~Engels$^{1}$}
\thanks{engels@wsu.edu}

\begin{abstract}
Controlling magnetism through non-magnetic means is highly desirable for future electronic devices, as such means typically have ultra-low power requirements and can provide coherent control. 
In recent years, great experimental progress has been made in the field of electrical manipulation of magnetism in numerous material systems. 
These studies generally do not consider the directionality of the applied non-magnetic potentials and/or magnetism switching. 
Here, we theoretically conceive and experimentally demonstrate a non-magnetic one-way spin switch device using a spin-orbit coupled Bose-Einstein condensate subjected to a moving spin-independent dipole potential. 
The physical foundation of this unidirectional device is based on the breakdown of Galilean invariance in the presence of spin-orbit coupling. 
Such a one-way spin switch opens an avenue for designing novel quantum devices with unique functionalities and may facilitate further experimental investigations of other one-way spintronic and atomtronic devices.
\end{abstract}

\affiliation{$^{1}$ Department of Physics and Astronomy, Washington State University,
Pullman, WA, USA 99164 \\
$^{2}$ Department of Physics, The University of Texas at Dallas, Dallas, TX
75080}
\maketitle

The ability to coherently control and switch the magnetism in a system plays a central role for building next-generation electronic devices, for example, magnetic memories and integrated circuits that rely on non-volatile information encoded in the direction of magnetization. 
Current technologies generally manipulate magnetism through methods involving magnetic fields or spin-polarized currents, such as spin-transfer torque (STT)\cite{SlonczewskiJC1996,BergerL1996}. 
Although the past decade has witnessed a remarkable development in STT-based spintronic devices\cite{STT}, switching the magnetism through non-magnetic means, such as electric fields, continues to be of high interest to significantly reduce the required switching power\cite{MatsukuraF2015}. 
To manipulate the magnetism, or overall spin, of a system with an electric field requires strong coupling between magnetic and electric properties and has been experimentally achieved recently in various materials including piezoelectric/multiferroic materials\cite{Eerenstein2006}, ferromagnetic semiconductors\cite{Yamada2011} and van der Waals magnets\cite{Jiang2018}.

\begin{figure}[t]
\includegraphics[width=0.45\textwidth]{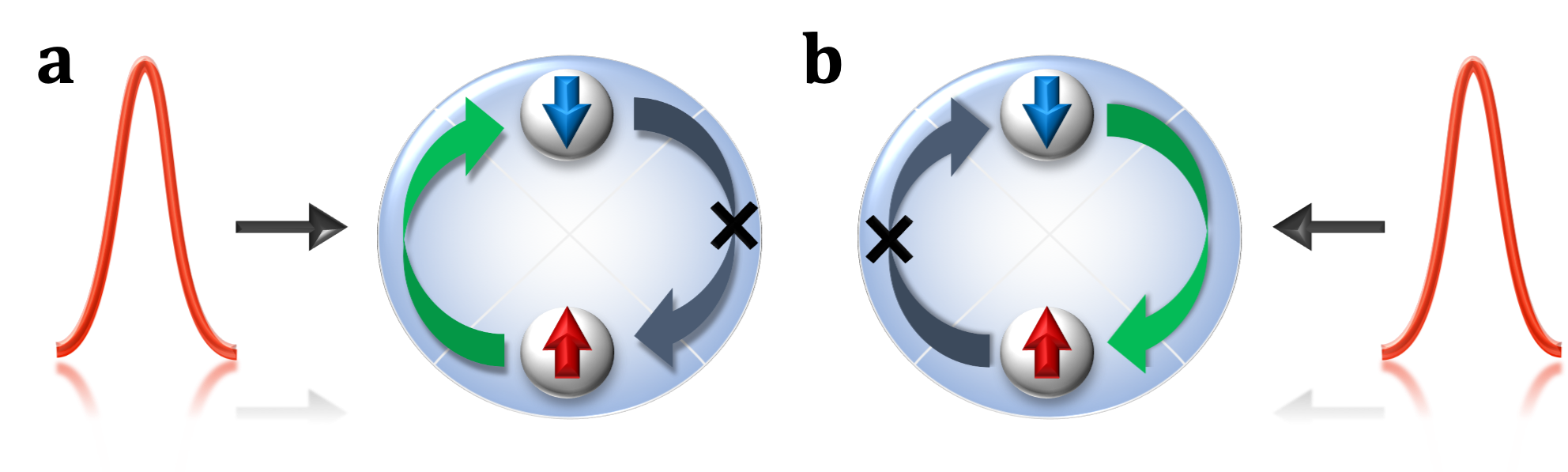}
\caption{
\textbf{Concept of a one-way non-magnetic spin switch.} A control signal entering from the \textbf{a} left, or \textbf{b} right, can switch the spin orientation from ($\uparrow$~to~$\downarrow$) or ($\downarrow$~to~$\uparrow$), while the reversed processes are forbidden.}
\label{fig1}
\end{figure}

Current non-magnetic spin switching devices are generally not spin-orientation selective; the spin switching can occur for both spin orientations ($\uparrow$ to $\downarrow$ and $\downarrow$ to $\uparrow$) and is insensitive to the orientation of external non-magnetic potentials.
Here, we introduce the concept of unidirectionality\cite{unidirection1,unidirection2,unidirection3} to spintronic devices and propose a non-magnetic one-way spin switch, whose basic concept is illustrated in Fig.~\ref{fig1}. 
When a non-magnetic control pulse interacts from the left (right), the spin orientation can be switched only from $\uparrow$ to $\downarrow$ ($\downarrow$ to $\uparrow$), while the reversed process is forbidden. 
Such a unique unidirectionality of a spin switch may greatly enhance our ability to manipulate magnetism for designing and engineering future spintronic devices.

One-way spin switching requires a strong coupling between the moving direction of the controlling potential and the spin flip in the device, which may naturally exist in a system with strong spin-orbit (\textit{i.e.}, spin-momentum) coupling. Spin-orbit (SO) coupling plays a crucial role for many condensed matter phenomena, including spintronics\cite{RMP-ST}. In this context, the experimental realization of SO coupling in ultracold atomic gases\cite{Lin2011,zhang2012collective,qu2013observation,olson2014tunable, wang2012spin,cheuk2012spin,wu2016realization,huang2016experimental,campbell2015itinerant,luo2016tunable,WilliamsRA2013,Lev2016,Ye2017,Fallani,Jo} provides a highly flexible and disorder-free platform for exploring spin-related quantum matter\cite{SOC1,SOC2,SOC3} and engineering atom-based spintronic devices\cite{Atomtronics}. Due to the coupling between momentum and spin, Galilean invariance is broken\cite{ZhuQ2012,ZhangYC2016}, indicating that the two momentum directions are no longer equivalent. The breakdown of Galilean invariance has recently been experimentally observed in a SO-coupled Bose-Einstein condensate (BEC) in an optical lattice\cite{HamnerC2015}.

In this work, we utilize the breakdown of Galilean invariance and experimentally realize the conceived non-magnetic one-way spin switch using a SO-coupled BEC (representing the device) subjected to a sweeping spin-independent Gaussian potential (representing the non-magnetic control signal). 
We observe that the efficiency of such a unidirectional spin switch strongly depends on the sweeping velocity of the potential and on the many-body interactions between atoms in the system. 
The experimental results agree well with numerical simulations based on mean field theory and can be explained through an intuitive two-step spin switch process mediated by the moving barrier and SO coupling.

\begin{figure}[t]
\includegraphics[width=0.48\textwidth]{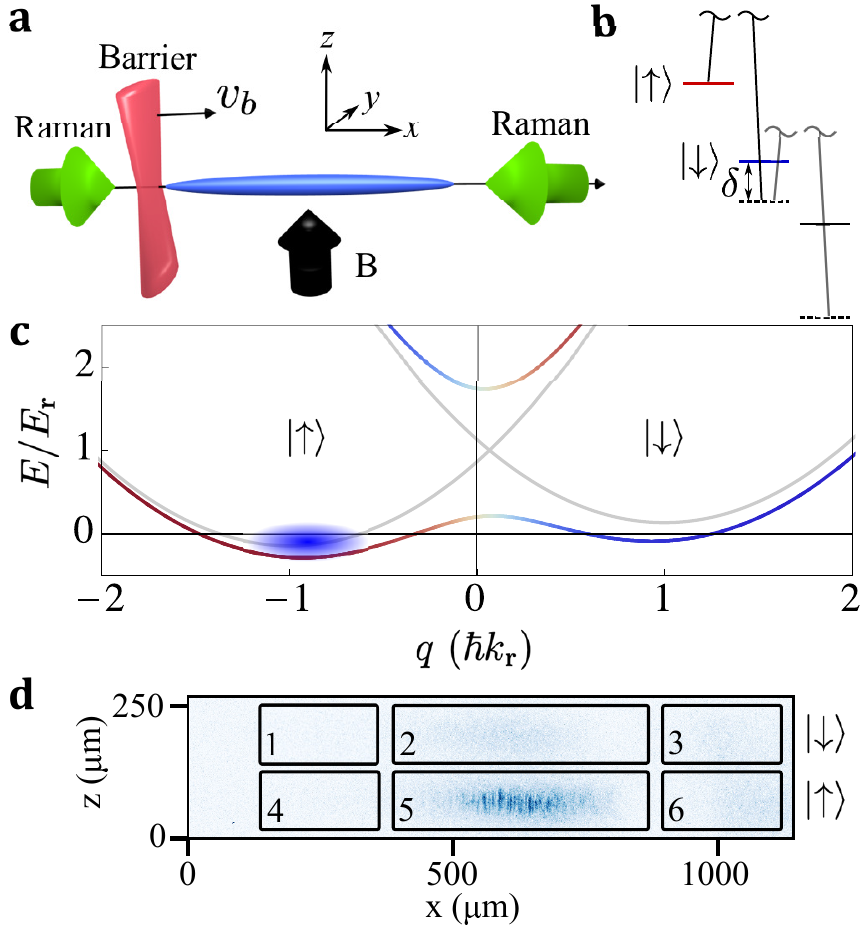}
\caption{
\textbf{Experimental setup.} 
\textbf{a} Two counter-propagating Raman beams applied along the elongated axis of the BEC induce SO coupling. A repulsive, spin-independent optical barrier is formed by a dipole sheet initially placed outside of the BEC and swept along the x-axis with a constant velocity $v_\mathrm{b}$. 
\textbf{b} Two pseudospin states, $m_\mathrm{F}=-1 (\equiv\:\mid\uparrow\rangle)$ and $0 (\equiv\:\mid\downarrow\rangle)$, are SO-coupled with coupling strength $\Omega$ and detuning $\protect\delta$, adjustable by the power and relative frequencies of the Raman beams. 
\textbf{c} The resulting SO-coupled double well band dispersion for parameters $\hbar\Omega =1.53E_\mathrm{r}$, $h\delta =0.27E_\mathrm{r}$. Here $E_\mathrm{r}$ is the atom recoil energy.  Colors of band represent spin composition of dressed states.
\textbf{d} Experimental image after the barrier has been swept through the BEC at $v_\mathrm{b}=+9$~mm s$^{-1}$. Numbered boxes indicate regions of interest sampled during analysis. Image procedure involves time-of-flight and Stern-Gerlach separation to split momentum and spin states, respectively.}
\label{fig2}
\end{figure}

\section{Results}
\subsection{Experimental realization of a unidirectional spin switch}

In the experiment, a $^{87}$Rb BEC is confined in an elongated optical dipole trap (Fig.~\ref{fig2}a,b) where two atomic hyperfine ground states $\mid \uparrow\rangle\equiv\:\mid\!F=1,m_\mathrm{F}=-1\rangle $ and $\mid\downarrow\rangle \equiv\:\mid\!F=1,m_\mathrm{F}=0\rangle $ are coupled through two counter-propagating Raman lasers, which induce SO coupling in the system (see Methods). 
The resulting dispersion features a double-well structure, as shown in Fig.~\ref{fig2}c.
With suitable laser detuning, $\delta $, atoms are prepared in the left well, with the majority of the atoms occupying the $\mid\uparrow\rangle$ state (Fig.~\ref{fig2}c). 
The control signal for the spin switch, \textit{i.e.}, the spin-independent potential barrier, sweeps through the BEC at a constant velocity, $v_\mathrm{b}$.

\begin{figure}[t]
\centering
\includegraphics[width=0.48\textwidth]{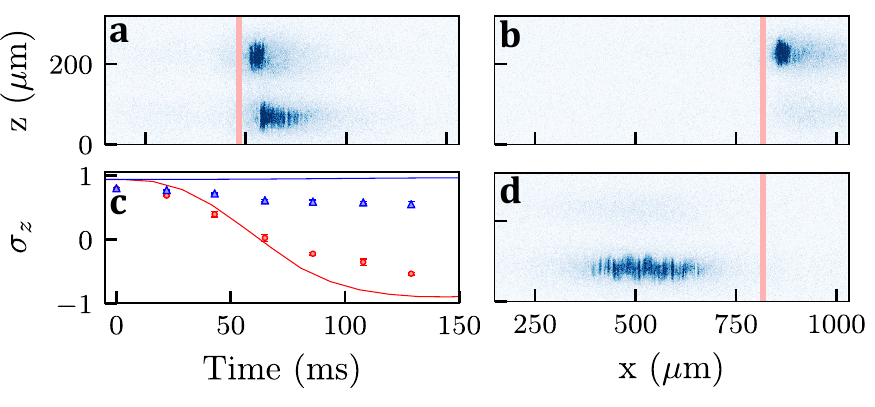}
\caption{
\textbf{Time evolution of spin switch.} 
Experimental image taken \textbf{a} at the middle ($t=65$~ms) and \textbf{b} at the end of a $v_\mathrm{b}=2~$mm s$^{-1}$ sweep. Red vertical lines indicate the position of the barrier just prior to imaging. 
\textbf{c} Evolution of spin polarization as $\pm 2$~mm s$^{-1}$ sweep progresses.
Data are mean $\pm $ s.d. for 5 experiments at each measured time for a barrier moving to the right (red circles) and left (blue triangles) at $v_\mathrm{b} = 2$~mm s$^{-1}$. 
Red and blue solid curves are corresponding numerical simulations for right and left moving
barriers, respectively. 
\textbf{d} Similar to \textbf{b} but with a $v_\mathrm{b}=15~$mm s$^{-1}$ sweep.}
\label{fig3}
\end{figure}

Absorption images, like those shown in Fig.~\ref{fig2}d and Fig.~\ref{fig3}a,b,d, are taken after the barrier has swept through the SO-coupled BEC. 
The evolution of the spin polarization, defined as $\sigma_z  = (N_\uparrow - N_\downarrow)/N_\mathrm{total}$, is  tracked with respect to time during a $v_\mathrm{b} = \pm 2$~mm s$^{-1}$ sweep in Fig.~\ref{fig3}c (see Supplementary Note 1).
Here, $N_{\uparrow }$ and $N_{\downarrow }$ are the atom numbers in spin $\mid \uparrow \rangle $ and $\mid \downarrow \rangle $ states respectively, and $N_{\text{total}}\ =N_{\uparrow }+N_{\downarrow }$.
At this featured velocity, $v_\mathrm{b}=+2$ mm s$^{-1}$ and for positive direction of motion (barrier moving in $+x$ direction), the BEC is reflected off the potential barrier and its spin polarization is gradually flipped from $\mid\uparrow\rangle$ to $\mid\downarrow\rangle$ (Figs.~\ref{fig3}a,b,c and Fig.~S1).
At large positive velocities, such as $v_\mathrm{b} = 15$~mm s$^{-1}$ (Fig. ~\ref{fig3}d), the BEC is fully transmitted through the potential barrier without a spin flip. 
A barrier moving in the opposite ($-x$) direction results in the overall spin of the system being nearly constant for all velocities, indicating that the barrier cannot switch the spin while traveling in this direction.

\begin{figure}[b]
\centering
\includegraphics[width=0.48\textwidth]{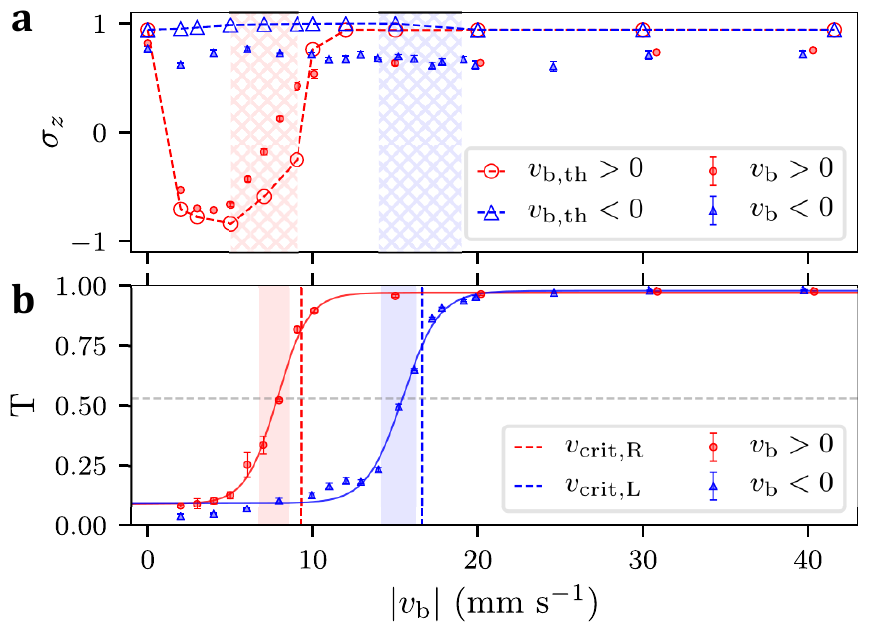}
\caption{
\textbf{Properties of unidirectional spin switch.} 
\textbf{a} Spin polarization versus $v_\mathrm{b}$ for positive (red circles) and negative (blue triangles) barrier velocities. 
Plotted data are mean $\pm $ s.d. for 5 experiments at each velocity. 
Hatched areas indicate experimentally observed heating. 
Numerical results for positive (red empty circles) and negative (blue empty triangles) barrier velocities are connected by dashed lines. 
\textbf{b} Transmission coefficient versus $v_\mathrm{b}$. 
Solid lines are sigmoidal fits to data, where shaded regions represent the 30\%-70\% error in the fit. 
Horizontal grey dashed line indicates crossover point of the fits at $T=0.53$. 
GPE numerical simulations result in sharp transitions between total reflection and total transmission (instead of a crossover region). The critical velocities for these transitions are represented by vertical dashed lines.}
\label{fig4}
\end{figure}

We measure the final spin-polarization, $\sigma _\mathrm{z}$, and the transmission coefficient, $T=N_\mathrm{trans}/N_\mathrm{total}$, after the barrier is swept through the full BEC for a large range of $v_\mathrm{b}$, where $N_\mathrm{trans}$ is the number of atoms behind the barrier. The experimental results are presented in Fig.~\ref{fig4}. 
We see three distinct regions for a barrier moving in the $+x$ direction (red data points): 

i) \textit{Low barrier velocity} ($0~\text{mm s}^{-1} < v_\mathrm{b}\lesssim 4$~mm s$^{-1}$). 
In this region, the atoms are strongly reflected from the moving barrier ($T\sim 0$). 
When the barrier velocity exceeds the lower critical velocity, $v_\mathrm{b} = 1.1~\text{mm s}^{-1}$ (see Theoretic Modeling), the spin polarization decreases, reaching a minimum at $v_\mathrm{b}\sim 4$~mm s$^{-1}$, where the spin switch efficiency, $(\sigma_{z,f}-\sigma_{z,i})/2$, is found experimentally to reach a maximum value of $85.8\pm0.85$\% ($\sigma_{z,\mathrm{min}} = -0.716\pm0.017$). 

ii) \textit{Medium barrier velocity} ($4~\text{mm s}^{-1}\lesssim v_\mathrm{b}\lesssim 11$~mm s$^{-1}$). 
In this region, the BEC is partially transmitted and partially reflected by the moving barrier. 
As the barrier velocity increases, the spin polarization increases, reaching a plateau around $v_\mathrm{b}\sim 11~$mm s$^{-1}$. This region where the BEC is transitioning between reflection and transmission is accompanied by heating in the experiment, depicted by hatched regions in Fig.~\ref{fig4}a. 
This crossover for a right moving barrier occurs at a critical velocity measured to be $7.819_{-0.764}^{+1.113} $~mm s$^{-1}$ using a sigmoidal best fit function (solid red line) in Fig.~\ref{fig4}b.

iii) \textit{High barrier velocity} ($v_\mathrm{b}\gtrsim 11~$ mm s$^{-1}$). 
In this region, the BEC is transmitted through the barrier without reflection ($T\sim 1$). 
The spin polarization plateaus at a constant value comparable to the initial spin polarization without a sweeping barrier ($t = 0$), indicating no spin switch occurs at high velocities. 

For a left moving barrier, three similar regions can be distinguished using the transmission coefficient, $T$.
No significant change of spin polarization is detected in this direction.
For a barrier traveling in the $-x$ direction, region (i) ends at $v_\mathrm{b}\sim -14$~mm s$^{-1}$ and region (iii) begins at $v_\mathrm{b}\lesssim-21$~mm s$^{-1}$. 
The critical velocity in region (ii) is described by a crossover experimentally determined to occur at $15.422_{-1.311}^{+0.845}$~mm s$^{-1}$, where heating is again observed within this region during the experiment. 
Since no significant change of spin polarization is observed for all negative velocities, these results show that our experimental setup successfully implements a non-magnetic one-way spin switch.

\begin{figure}[tbp]
\centering
\includegraphics[width=0.48\textwidth]{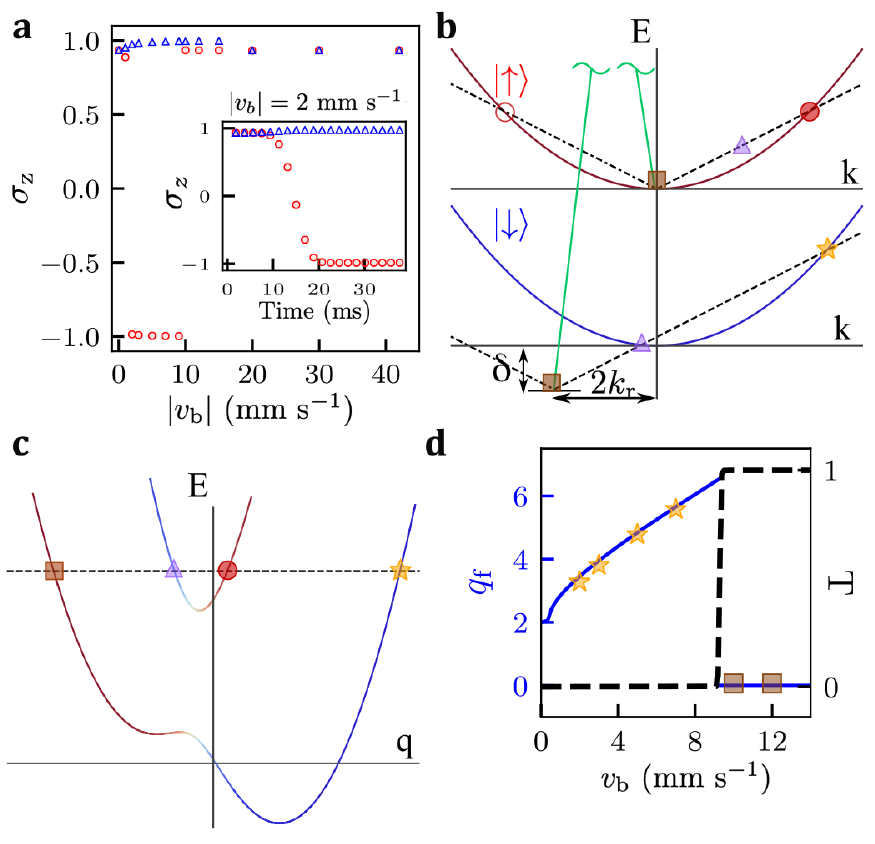}
\caption{
\textbf{Unidirectional spin switch mechanism.} 
\textbf{a} Noninteracting GPE simulation of spin polarization versus $v_\mathrm{b}$ (interacting case in Fig.~\ref{fig4}a).
The inset shows the evolution of spin polarization during a highly efficient $v_\mathrm{b}=2$~mm s$^{-1}$ sweep (corresponding to the interacting case in Fig.~\ref{fig3}c). 
Illustration of the coupling channel mechanism in the \textbf{b} laboratory frame and in the \textbf{c} moving frame with positive (black dashed line) or negative (black dot-dashed line) barrier velocity.
Atoms are prepared in the $\mid\uparrow\rangle$ spin state and coupled to the $\mid\downarrow\rangle$ spin state by means of Raman coupling (solid green lines). 
Five channels emerge: $C_\mathrm{T_1}$(brown squares), $C_\mathrm{R_1}$ (red filled circles), $C_\mathrm{T_2}$ (purple triangles), $C_\mathrm{R_2}$ (orange stars), and $C_\mathrm{R_L}$ (open red circle).
\textbf{d} Final quasimomentum $q_\mathrm{f}$ (solid blue curves), and transmission coefficient T (black dashed), calculated from theory for a barrier with $v_\mathrm{b}>0$. 
Results from GPE simulations are correspondingly labelled.}
\label{fig5}
\end{figure}

\subsection{Theoretic modeling}

To elucidate the physical mechanisms of the non-magnetic one-way spin switch, we perform simulations of the Gross-Pitaevskii equation (GPE) that describes the dynamics of the BEC under the mean field approximation.
These results are included in Fig.~\ref{fig3}c and Fig.~\ref{fig4}, and are in good agreement with experimental observations. 
The fact that the experimentally determined spin polarizations for large velocities lie below the GPE results is due to background contamination of the absorption images from residual thermal atoms created by heating from the Raman beams.
As illustrated by the GPE results in Fig.~\ref{fig4}a, the spin switch efficiency does not reach 100\%.
This is mainly attributed to many-body interaction between atoms in the system where, during the spin switch process, the coexistence of opposite spins leads to a strong density modulation in the BEC.
This effect results in an increase in the density-density interaction energy, preventing an perfectly efficient spin flip in the system (See Supplementary Note 2). 
This explanation is corroborated by numerical results for noninteracting atoms shown in Fig.~\ref{fig5}a and Fig.~S2.  
In this case, the spin switch mechanism can indeed reach an efficiency close to 100\%. 
To understand the underlying physical mechanism driving the spin switch in our system, the single particle (noninteracting) picture is used in the discussion below.

The spin switch relies on momentum and energy transfer during two simultaneous processes: a kick imparted by a moving barrier and a Raman transition. 
In the laboratory frame, the moving Gaussian barrier depends on both time and space, with a Fourier
spectrum
\begin{equation}
\tilde{V_\mathrm{b}}(\Delta k,\omega )=\digamma\!\left( \Delta k\right) \delta (2 \Delta k v_\mathrm{b}/v_\mathrm{r}-\omega),
\end{equation}
where $\delta (x)$ is the Dirac delta function, $v_\mathrm{r}= \hbar k_\mathrm{r} / m$ is the recoil velocity, and 
\begin{equation*}
\digamma\! \left( \Delta k\right) =U_\mathrm{b}\sqrt{\pi/2w_\mathrm{b}^{2}}\exp \left( -\left( \Delta k\right) ^{2}w_\mathrm{b}^{2}/8-i\Delta kx_{0})\right)
\end{equation*} 
is a function used to describe the Gaussian barrier where $U_\mathrm{b}$, $w_\mathrm{b}$ and $x_{0}$ are barrier height, width, and initial position, respectively. 
Here, and in the following discussion, the recoil momentum, $k_\mathrm{r} = 2\pi/\lambda_\mathrm{r}$, and recoil energy, $E_\mathrm{r}=(\hbar^2 k_\mathrm{r}^2) / 2m$, are chosen as units. 

The delta function forces the momentum and energy transfer induced by the barrier to satisfy $\omega =2 \Delta k v_\mathrm{b}/v_\mathrm{r}$. 
From this stipulation, the conservation of energy and momentum lead to resonance conditions $E_\mathrm{i}=E_\mathrm{f}-\omega $ and $q_\mathrm{i}=q_\mathrm{f}-\Delta k$, where $E_\mathrm{i,f}$ ($q_\mathrm{i,f}$) are the initial and final atomic energy (quasimomentum). 
In the SO-coupled picture, quasimomentum is related to kinetic momentum through $q_{\mathrm{s}\sigma}=k_{\mathrm{s}\sigma }\mp 1$, where $-$ ($+$) is for spin $\mid\uparrow\rangle$ ($\mid\downarrow\rangle$), $\mathrm{s}$ represents the state (either i or f), and $\sigma$ is the $\uparrow$ or $\downarrow$ spin state.

As illustrated in Fig.~\ref{fig5}b, for atoms initialized in the $\mid\uparrow\rangle$ state at the band minimum ($k_\mathrm{i\uparrow }=0$, $E_\mathrm{i}=0$), the right-moving barrier ($v_\mathrm{b}>0$) drives atoms along the black dashed line (with a slope $2 v_\mathrm{b}/v_\mathrm{r}$). 
In the absence of Raman coupling, two resonance channels $C_\mathrm{T_1}$ (brown square) and $C_\mathrm{R_1}$ (red filled circle) exist within the upper spin state.
These channels do not entail a spin flip and are solutions of the resonance conditions $k_{f\uparrow }=\Delta k$, $\omega =2 \Delta k v_\mathrm{b}/v_\mathrm{r}=E_{f}\equiv k_{f\uparrow }^{2}$, yielding $\Delta k=2 v_\mathrm{b}/v_\mathrm{r}$. 
Raman coupling in the system (green solid lines) induces two additional resonance channels $C_\mathrm{T_2}$ (purple triangle) and $C_\mathrm{R_2}$ (orange star) which involve the coupled, lower spin state and thus entail a spin flip. 
The resonance conditions in the presence of SO coupling and a moving Gaussian barrier become $k_{f\downarrow }=\Delta k-2$, $\omega =2 \Delta k v_\mathrm{b}/v_\mathrm{r}=E_{f}\equiv k_{f\downarrow }^{2}+\delta$, where $\delta$ is the detuning parameter of the SO-coupled system.

A similar analysis applies in the co-moving barrier frame with the quasimomentum band structure and resonance channels shown in Fig.~\ref{fig5}c for positive velocities (see Supplementary Note 3). 
For large barrier velocities (regions ii and iii from the discussion above), the resonance channels $C_\mathrm{R_1}$ and $C_\mathrm{T_2}$ cannot be accessed due to an avoided band crossing (see Fig.~\ref{fig5}c).
Instead, atoms in these regions prefer the transmission (reflection) channel $C_\mathrm{T_1}$ ($C_\mathrm{R_2}$) if the barrier velocity is larger (smaller) than a critical value, $v_{\mathrm{crit}}^{\mathrm{R}}$. 
This is confirmed by numerical simulations shown in Figs.~5A and~5D. 
As the velocity decreases (within region i), the system switches from $C_\mathrm{R_2}$ to $C_\mathrm{R_1}$, as a low velocity barrier cannot drive atoms over the band-barrier around $q=0$ in the double well band dispersion (see Fig.~\ref{fig2}c and Fig.~S4).

The left-moving barrier drives the atoms along the black dot-dashed line in Fig.~\ref{fig5}b with a slope $-2|v_\mathrm{b}|/v_\mathrm{r}$. 
There are only two resonance channels in this case, $C_\mathrm{T_1}$ (brown square) and $C_\mathrm{R_L}$ (red open circle), both of which result in no spin flip. 
The system would therefore prefer the transmission (reflection) channel $C_\mathrm{T_1}$ ($C_\mathrm{R_L}$) if the barrier velocity is larger (smaller) than a critical value $v_{\mathrm{crit}}^{\mathrm{L}}$. 
The Raman process is no longer involved for a left-moving barrier, in contrast to the Raman assisted reflection channel $C_\mathrm{R_2}$ for a right-moving barrier. 
The momentum kick induced by the Raman coupling leads to a smaller critical velocity for a right-moving barrier ( $v_{\mathrm{crit}}^{\mathrm{R}}<v_{\mathrm{crit}}^{\mathrm{L}}$). 
A direct theoretical calculation gives $v_{\mathrm{crit}}^{\mathrm{L}}=16.8$~mm s$^{-1}$ and $v_{\mathrm{crit}}^{\mathrm{R}}=9.3$~mm s$^{-1}$ (see Supplementary Note 3), which are consistent with experiments (see Fig.~\ref{fig4}b).

\section{Discussion}

The coupling channels developed here under the single particle picture continue to exist even when interactions are introduced into the GPE simulations. 
Therefore the spin-flip dynamics are still observable in experiments.
However, as the interaction strength increases, intense density and spin modulations in real space and various excitations in momentum space emerge, deteriorating the near perfect efficiency of the coupling channels. 
In this context, the spin switch efficiency could be further improved by reducing the atomic interaction through smaller atomic number, weaker confinement along the radial direction, Feshbach resonances, etc. 
Since the spin switch mechanism relies on single particle physics, it may also apply to SO-coupled degenerate Fermi gases, so long as the initial momentum and energy distributions of the atoms satisfy the energy and momentum resonance conditions. 
Similar unidirectional spin switches may also be engineered in electronic materials, for instance, in spin-orbit coupled nanowires with electrically-controlled moving potential pulses. 
For future studies, it would be intriguing to investigate the one-way spin switch in a 2D SO-coupled system (for example, in a system subjected to Rashba coupling) that has been realized in experiments for both Bose and Fermi gases\cite{huang2016experimental,wu2016realization}. 
In summary, our proposed concept of a unidirectional spin switch device and its experimental realization opens an avenue for designing innovative quantum devices with unique functionalities and may facilitate further experimental investigations of other one-way spintronic and atomtronic devices.

\section{Methods}
\begin{small}
\noindent\textbf{Experimental Preparation}
Our experiments begin with $7\times 10^{5}$ $^{87}$Rb atoms in a BEC held in an elongated optical dipole trap with trap frequencies $\{\omega _{x},\omega _{y},\omega _{z}\}=2\pi \times \{3.07,278,278\}~\text{Hz}$. The longitudinal speed of sound in the center of the BEC is $c_\mathrm{s}\approx 2.2$~mm s$^{-1}$. A 10~G magnetic field in the z-direction splits the internal $|F=1\rangle$ hyperfine states according to the Zeeman shift. The atoms are initially prepared in the $|F,m_\mathrm{F}\rangle=|1,-1\rangle $ ground state.

Two counter-propagating Raman beams ($\lambda _\mathrm{r}=789.1$~nm) are applied along the x-direction. 
The Raman beams are first ramped on, with large Raman detuning, to a Raman coupling strength of $\hbar \Omega =1.53~E_\mathrm{r}$, where the recoil energy $E_\mathrm{r}=(\hbar k_\mathrm{r})^{2}/2m=h\cdot 3684 $~Hz and the recoil momentum $k_\mathrm{r}=2\pi /\lambda _\mathrm{r}$. 
The detuning of the Raman drive is then adiabatically reduced to a final value of $\delta =0.27~E_\mathrm{r}=1\pm0.030$~kHz in 100~ms.
This adiabatic loading ``dresses'' the atoms with SO coupling of the above parameters. 
The Raman beams coherently couple the $|1,-1\rangle $ and the $|1,0\rangle $ states, while the $|1,+1\rangle $ state is decoupled due to the quadratic Zeeman shift. 
This generates an effective pseudospin-1/2 system with $\mid\uparrow\rangle \equiv |1,-1\rangle $ and $\mid\downarrow\rangle \equiv |1,0\rangle$.

While preparing the atoms, a repulsive barrier is jumped on either to the left or right of the BEC (see Fig.~\ref{fig2}a).
This barrier is generated by a laser propagating in the $z$-direction, with wavelength $\lambda _\mathrm{b}=660$~nm, and Gaussian widths of $\{w_{x},w_{y}\}=\{11,63\}~\mu $m. 
This laser creates a repulsive potential of height $U_\mathrm{b}=8.3E_\mathrm{r}\approx 15\mu $, where $\mu$ is the chemical potential at the center of the BEC. 
Once the atoms are dressed in the the SO-coupled $\mid\uparrow\rangle$ state, a galvanometer is then used to sweep the barrier through the BEC along the \textit{x}-direction at different constant velocities, $v_\mathrm{b}$. 

After the sweep, all optical fields are jumped off and absorption imaging is performed along the \textit{y}-direction after 10.1~ms time-of-flight expansion, during which a Stern-Gerlach technique is used to spatially separate the two spin components. 
For this experimental setup, sweeping in the $+x$ ($-x$) direction is associated with the $+q$($-q$) direction in the SO-coupled dispersion.

The spin polarization, $\sigma _{z}=(N_{\uparrow }-N_{\downarrow })/N_{\text{total}}$, and the transmission coefficient, $T=N_{\text{trans}}/N_{\text{total}}$, are measured after the barrier has passed through the BEC. 
The number of atoms in the $\mid\downarrow\rangle$ or $\mid\uparrow\rangle$ state are defined as $N_{\downarrow,\uparrow}=\sum_{i=1,4}^{3,6} N_{i}$, with the total atom number, $N_{\text{total}}\ =N_{\uparrow }+N_{\downarrow }$ and numerical subscripts refer to the numbered regions in Fig.~\ref{fig2}d of the main text. 
The number of atoms transmitted through the barrier during the sweep is defined depending on the sweep direction by $N_{\text{trans}}\ =N_{2}+N_{2\pm1}+N_{5}+N_{5\pm1}$, where $- (+)$ is applied for right (left) sweeping barriers. 

In addition to data where the barrier is swept through the BEC, data was recorded where no sweep occurred to understand heating effects due to the presence of the Raman beams over time.
Fig.~\ref{fig4}a of the main text shows an initial spin polarization (at $t=0$) of $\sigma_{z,0\mathrm{ms}} = 0.818\pm0.015$.
As the atoms are held in trap in the presence of SO coupling, the Raman beams induce heating. 
The longest time the atoms are held in the dressed state, corresponding to a $v_\mathrm{b}=2$~mm s$^{-1}$ sweep time, is 258~ms. 
Holding the atoms for this amount of time results in a reduced spin polarization of $\sigma_{z,258\mathrm{ms}} = 0.450\pm0.044$.  
The discrepancy between theoretical prediction and experiments is thus attributed to background contamination of the images from residual thermal atoms created by heating from the Raman beams.
For example, a 10\% thermal background population with respect to the total number of atoms in the system results in the spin polarization being reduced from $\sigma_{z}~\mathrm{to}~0.82\sigma_{z}$.

\vspace{4mm}\noindent\textbf{GPE numerics}
The experimental system and its dynamics are described by a dimensionless Gross-Piteavskii equation (GPE),
\begin{equation}
i\frac{\partial }{\partial t}\Psi =\left( H_{0}(x)+V_\mathrm{b}(x,t)+\frac{g}{2}|\Psi |^{2}\right) \Psi,
\end{equation}%
under the mean-field approximation, where $\Psi =(\psi_{\downarrow },~\psi _{\uparrow })^{T}$ is the two-component wave function.
The interaction strength between the two spin components in the system is given by $g_{\downarrow \downarrow}=g_{\downarrow\uparrow}=g_{\uparrow \downarrow}=0.9954g_{\uparrow \uparrow}$ for $^{87}$Rb, with the corresponding effective 1D interaction strength $g_{\uparrow \uparrow }=1426$, due to wavefunction normalization in the simulations. 
The time-independent, single particle Hamiltonian for particles dressed with SO coupling is given by
\begin{equation}
H_{0}(x)=\left(
\begin{array}{cc}
p_{x}^{2}-\frac{\delta }{2} & \frac{\Omega }{2}e^{-i2x} \\
\frac{\Omega }{2}e^{i2x} & p_{x}^{2}+\frac{\delta }{2}
\end{array}
\right) +\frac{1}{2}\omega _{x}x^{2},
\end{equation}
where $\omega _{x}$ is the trapping frequency along $x$. 
Finally, the space and time-dependent potential is described by an optical Gaussian beam,
\begin{equation}
V_\mathrm{b}(x,t)=U_\mathrm{b}\exp \left( -2(x-(x_{0}-v_\mathrm{b}t))^{2}/w_\mathrm{b}^{2}\right),
\end{equation}
where $U_\mathrm{b}$ is the height, $x_{0}$ is the initial position, $v_\mathrm{b}$ is the sweeping speed, and $w_\mathrm{b}=w_{x}$ denotes the barrier width. 
The recoil energy, $E_\mathrm{r}$ (recoil momentum, $k_\mathrm{r}$), are set as the unit for energy (momentum) in the system.

\end{small}

\vspace{2mm} {\noindent\textbf{Acknowledgements}
\newline
M.E.M. and P.E. acknowledge support from the NSF (Grant No. PHY-1306662).
J.H., X.-W.L. and C.Z. acknowledge support from AFOSR (FA9550-16-1-0387),
NSF (PHY-1806227), and ARO (W911NF-17-1-0128). 

\vspace{2mm} \noindent\textbf{Author contributions} 
\newline
M.E.M and P.E. performed apparatus preparation, data collection, data
analysis, and contributed to the writing of the manuscript. J. H, X.-W.L.,
and C.Z performed GPE numerics, system modeling, and contributed to the
writing of the manuscript. C.Z. and P.E. supervised this project. 

\vspace{2mm}\noindent\textbf{Competing interests} 
\newline
The authors declare no financial or non-financial competing interests.

\vspace{2mm} \noindent\textbf{Data availability} 
\newline
All relevant experimental and numerical datasets in this work will be made available from the corresponding authors upon reasonable request.


\pagebreak
\widetext
\begin{center}
\textbf{\large Supplemental Material}
\end{center}
%
\renewcommand{\theequation}{\arabic{equation}}
\setcounter{equation}{0}
\renewcommand{\figurename}{Supplementary Figure}
\setcounter{figure}{0}
\setcounter{page}{1}
\makeatletter

\subsection{Supplementary Note 1}
\label{sec:spin-flip-evolution}

To observe the spin-flip evolution, sequential images of the SO-coupled BEC are taken during the sweep. 
Instead of completing a full sweep, the barrier is stopped after a time $t_\mathrm{sweep}$ at which $\sigma _{z}$ is analyzed. 
Supplemental Fig.~\ref{figs1}a-d shows sequential images of spin-flip dynamics for a barrier moving at $v_\mathrm{b}=2$~mm s$^{-1}$. 
In this sequence, $t_\mathrm{sweep}=98$~ms corresponds to a sweep that ends at the center of the BEC. 

In panel a, the barrier just reaches the left edge of the BEC, and the condensate remains in the $\mid\uparrow\rangle$ state. 
As the non-magnetic barrier starts to interact with the atoms, we observe a portion of the $\mid \uparrow \rangle$ state atoms being transferred to the $\mid\downarrow\rangle$ state. 
At this barrier velocity, the BEC is totally reflected by the Gaussian potential. 
The spin-flip process continues as the barrier continues to interact with and drive through the BEC as shown in panel c. 
At the end of the sweep, nearly all atoms occupy the $\mid\downarrow\rangle$ state (panel d) and the real-space profile of this component is non-Gaussian in shape. 

Corresponding GPE numerical solutions are shown in Supplemental Fig.~\ref{figs1}e-h for sweeping velocity $v_\mathrm{b}=2$~mm s$^{-1}$.
The 1D profiles provide density distributions for each spin component along the sweeping direction. 
We find good agreement here between experimental and simulated GPE results. 

\begin{figure}[tbph]
\centering\includegraphics[width=0.7\textwidth]{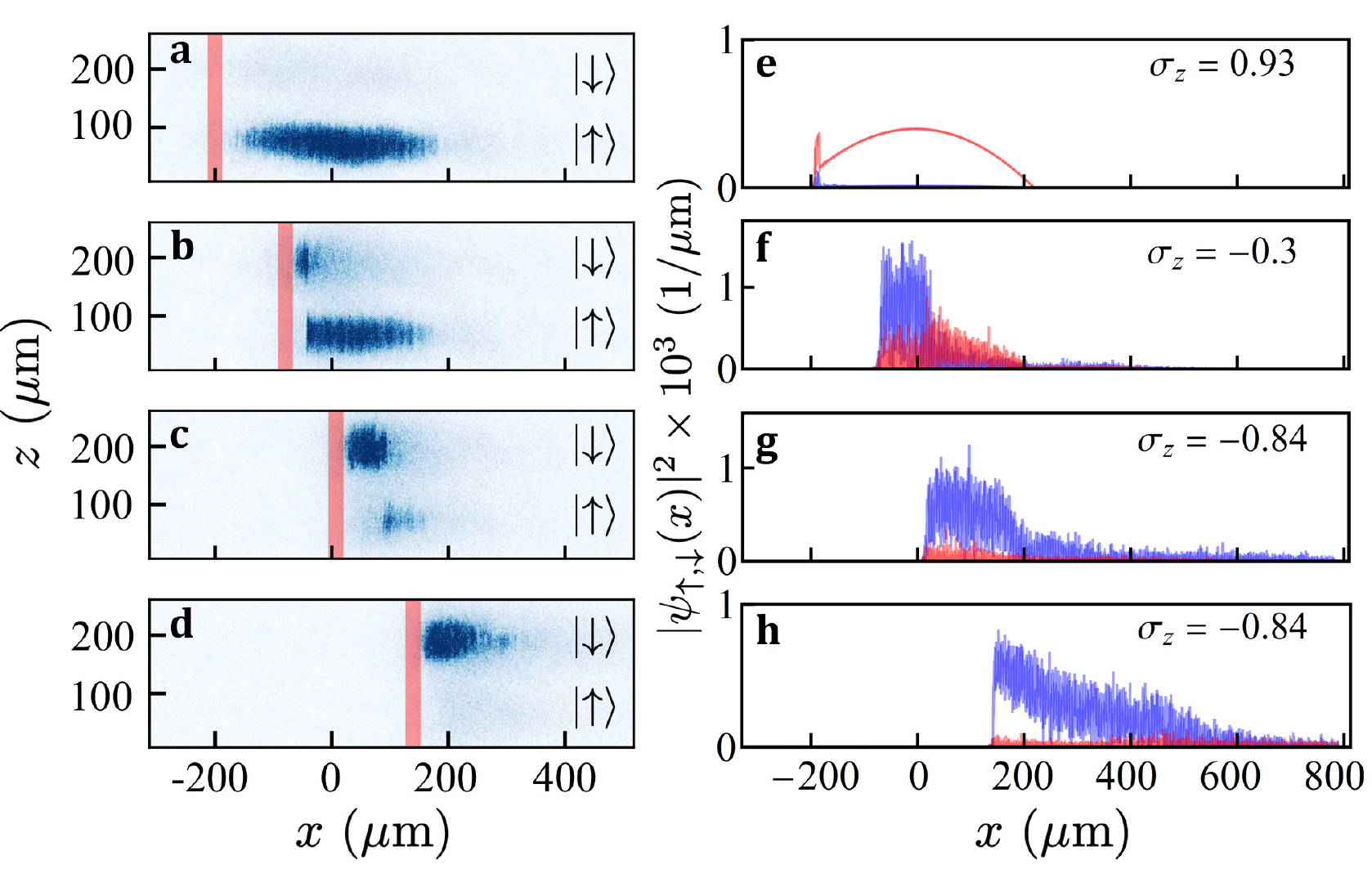}
\caption{
\textbf{Spin flip evolution.} 
Barrier moves at $v_\mathrm{b}=2$~mm s$^{-1}$ to the right.  Atoms originally in the $\mid\uparrow\rangle$ spin state are transferred to the $\mid\downarrow\rangle$ spin state as the sweep progresses. 
\textbf{a-d} Experimental images taken at a time, $t$, into the sweep. Red vertical line represents the position of the barrier prior to imaging. Spin states are separated during imaging using a Stern-Gerlach technique. SO coupling parameters are the same as those in Fig.~2c of the main text.
\textbf{e-h} Corresponding integrated cross sections from numerical GPE simulations. Red (blue) represents populations the $\mid \uparrow \rangle $ ($\mid \downarrow \rangle$) spin state. Spin polarizations are calculated from GPE simulations found in each panel.
Corresponding experimental and numerical images are taken after \textbf{a,e} $t=22$~ms, \textbf{b,f} $t=87$~ms, \textbf{c,g} $t=130$~ms, and \textbf{d,h} $t=195$~ms into the sweep.}
\label{figs1}
\end{figure}

\subsection{Supplementary Note 2}
\label{sec:Sim-momentum-space-profile}

Numerical simulations are conducted for a range of sweeping velocities, as presented in Supplemental Fig.~\ref{figs2}a. 
The results, highlighted in Fig.~4 of the main text, are consistent with experimental observations and have been discussed in the main text.
For speeds at which the BEC is fully transmitted through the barrier, the change in the momentum distribution is  small. 
As the speed of the barrier decreases and an increasingly larger fraction of the condensate is reflected by the moving barrier, a spin flip process occurs for positive velocities, despite the spin independence of the potential. 
Momentum-space analysis reveals that the momentum peak of the resulting BEC is shifted and split for slow barrier velocities, leading to rapid density modulations in real space due to the superposition of different plane wave states. 
For example, when $v_\mathrm{b}=7$~mm s$^{-1}$, the momentum of the spin-flipped component is shifted to $\sim 2.2~k_\mathrm{r} \pm 0.5~k_\mathrm{r}$.

Turning off the interatomic interactions, i.e. letting $g = 0$, we can elucidate the role of the interactions in order to further understand how large momentum states are excited in our system.
Numerical results for the single particle case are plotted in Supplemental Fig.~\ref{figs2}b and can be compared with the interacting case in Supplemental Fig.~\ref{figs2}a. 
Similar to the interacting case, fast barrier sweeps ($v_\mathrm{b}>9.3$~mm s$^{-1}$) result in a nearly unperturbed final state, while slow barrier sweeps result in total reflection from the barrier with a spin flip when sweeping in the positive direction. 
For example, when $v_\mathrm{b}=7$~mm s$^{-1}$, the spin is flipped with $\sim$100\% efficiency and the condensate is kicked to a larger momentum state than in the case with interactions, $k\approx 3.7~k_\mathrm{r}$. 
An important observation is that in the single particle case, the density profiles in real- and momentum-space are smooth with ideal Gaussian-like distributions. 
This is due to the sharp spin switch transition with $\sim$100\% efficiency, instead of the smooth crossover for the spin polarization in the interacting regions, which involves the superposition of multiple momentum states for the final state.

When decreasing the barrier velocity to $\sim1$~mm s$^{-1}$, no spin flip occurs even for the positive sweep direction. 
The BEC is still completely reflected. 
In this case, the momentum peak is shifted to $k=0.35~k_\mathrm{r}$, which is $\sim 2\times$ the sweeping speed in units of $k_\mathrm{r}$. 
This indicates a simple single particle reflection process that is due to the suppression of the spin-flip reflection channel ($C_\mathrm{R_2}$) at small velocities and can be predicted in the co-moving frame discussed in the Note \ref{sec:scatt-dynamics}. 
Experimentally, this regime is hard to observe because low velocities require long sweep times, leading to noticeable atom loss from the BEC due to Raman-induced heating. 

In the presence of a barrier sweep, the change in momentum can be exactly predicted using the coupling-channel theory resonance conditions developed in the main text. 
Special emphasis for this work has been put on the reflection channel, $C_\mathrm{R_2}$, because it leads to a spin-flip in our experimental system. 
Another spin-flip channel, $C_\mathrm{T_2}$, exists but is suppressed by the avoided band crossing in the system.
This can be overcome through Landau-Zener tunneling, but that is an undesirable feature for a functional spin switch. 
A numerical study of this is presented in Note \ref{sec:barrier-width-effects}.

\begin{figure}[t]
\centering\includegraphics[width=1\textwidth]{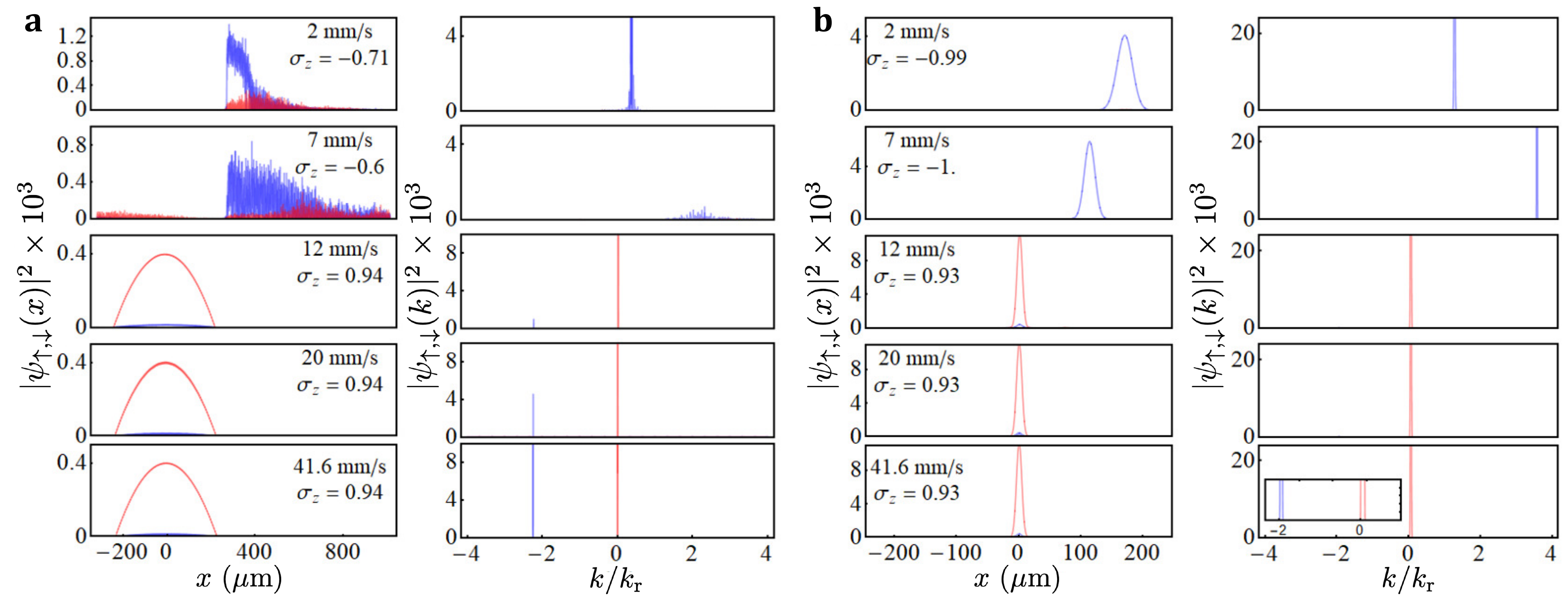}
\caption{\textbf{Simulations with positive velocities.} 
\textbf{a} GPE simulations for experiments at different (positive) sweeping velocities. 
Panels show the final real- and momentum-space distributions after the sweep. 
\textbf{a} Similar to \textbf{a} but in the single particle regime ($g=0$). 
The inset shows a zoomed-in view  in the momentum-space distributions.}
\label{figs2}
\end{figure}

\subsection{Supplementary Note 3}
\label{sec:scatt-dynamics}

In Fig.~4a of the main text, we observe a spin flip in the region of low, right moving barrier velocities where, below the critical velocity, the spin polarization has a smooth dependence on velocity.
The corresponding region for the noninteracting case in Fig.~5c has a spin polarization of $\sigma _{z}\approx -1$ and exhibits a discontinuous jump to $\sigma_z \approx +1$ at a critical velocity. 
The critical velocities where $\sigma _{z}$ changes from negative to positive values are similar in both cases, indicating that the origin of the spin-flip process can be understood by analyzing the single-particle dynamics (i.e. the non-interacting case). 
Here, we focus on the transmission coefficient, as the coupling channels are investigated in the main text. 

\begin{figure}[tbp]
\includegraphics[width=0.6\textwidth]{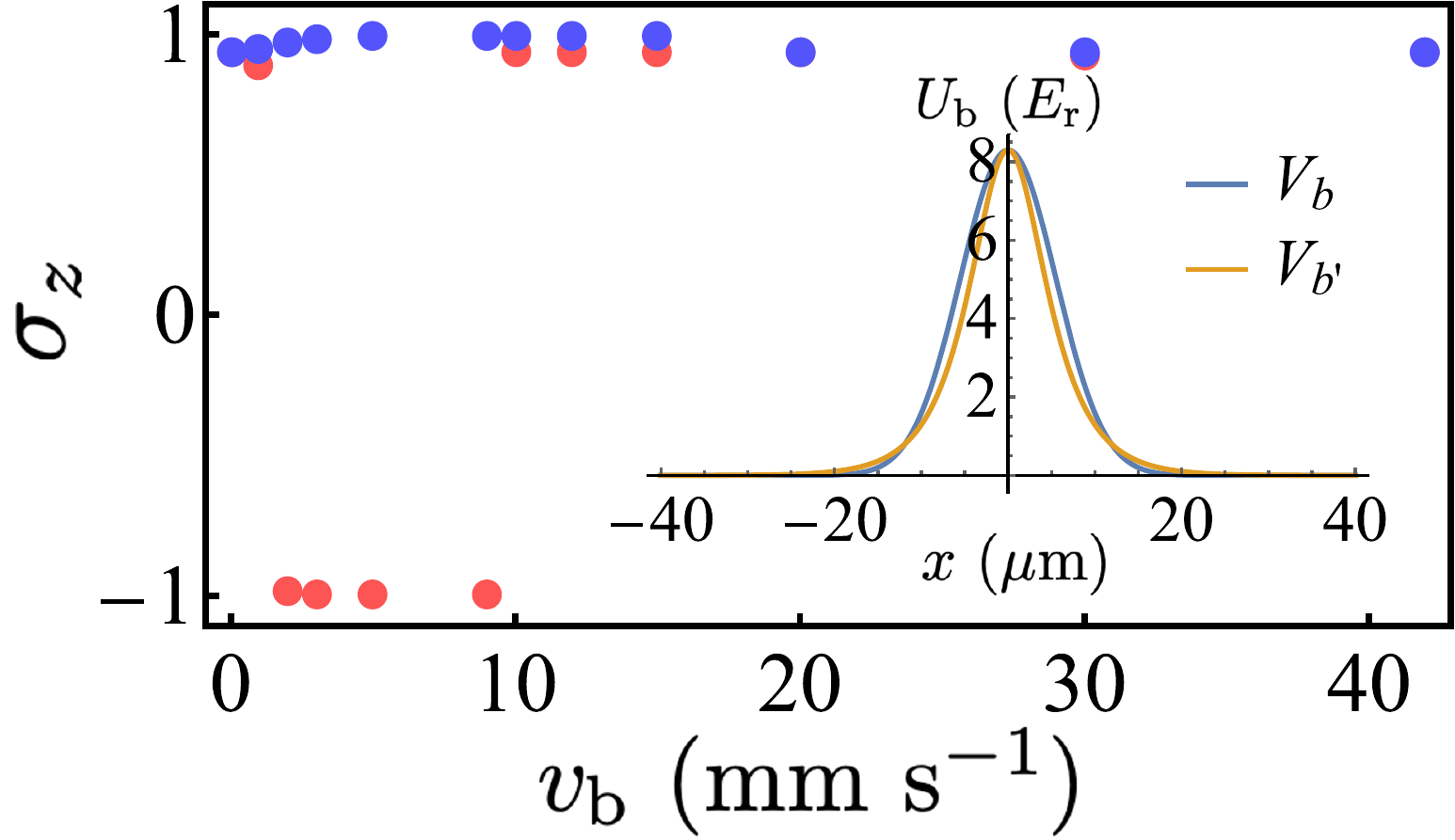}
\caption{
\textbf{Non-interacting GPE simulations} The dependence of the final polarization on
the sweeping speed for a single particle with the sweeping barrier $V_\mathrm{b^{\prime }}$ instead of a Gaussian barrier. Inset shows the potential
profile for each barrier. }
\label{figs3}
\end{figure}

With no known analytic solution for the scattering of a single particle by a Gaussian barrier, we consider a potential with a similar real space profile (see Supplemental Fig.~\ref{figs3} inset), 
\begin{equation}
V_\mathrm{b^{\prime}}(x)=U_\mathrm{b^{\prime }}/\cosh ^{2}(x/w_\mathrm{b^{\prime }}),
\end{equation}
where $U_\mathrm{b^{\prime }}$ and $w_\mathrm{b^{\prime }}$ are the potential height and width of the profile.
This potential yields nearly identical dynamics in GPE simulations, as shown in Supplemental Fig.~\ref{figs3}. 
The scattering process for this potential is characterized by a single dimensionless parameter $\nu =w_\mathrm{b^{\prime }}^{2}U_\mathrm{b^{\prime }}/2\gg 1>1/8$ and the transmission coefficient $T$ is given by\cite{HaarD1975}
\begin{equation}
T=\frac{\sinh ^{2}(\pi \sqrt{2\epsilon })}{\sinh ^{2}(\pi \sqrt{2\epsilon })+\cosh ^{2}(\frac{\pi }{2}\sqrt{8\nu -1})},
\end{equation}
where $\epsilon $ is the single-particle energy. 
This expression is obtained without SO coupling. 
The resulting critical speed, $v_\mathrm{crit}$, for the transition from total reflection to total transmission is $v_\mathrm{b} = -2.85v_\mathrm{r} \approx -16.6$~mm s$^{-1}$, relevant for a left sweeping barrier in our system. 
For a barrier moving to the right, our numerics show that the critical velocity in the presence of SO coupling is reduced to $v_\mathrm{b} = 1.6v_\mathrm{r}\approx 9.3$~mm s$^{-1}$. 
These values are obtained in the single particle regime ($g=0$). 
As shown in Fig.~4b of the main text, even for the interacting system the transition region is relatively narrow due to large $\nu $, corresponding to the high and wide barrier in experiments. 
The predictions for the critical velocities are consistent with both GPE simulation and experimental measurements (see Supplemental Fig.~\ref{figs4}a,b and Fig.~4b in the main text).

\begin{figure}[tbp]
\includegraphics[width=0.7\textwidth]{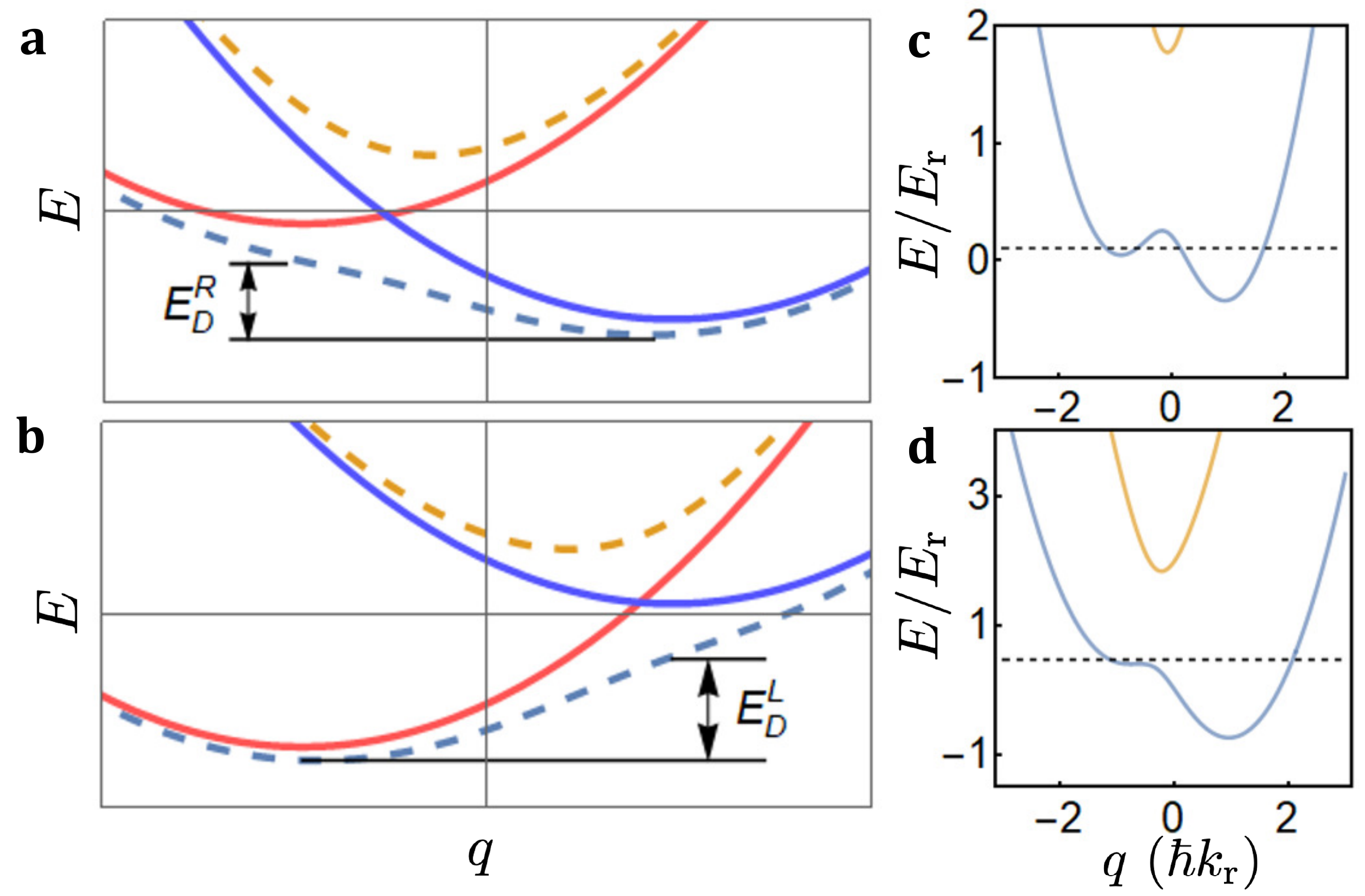}
\caption{
\textbf{Co-moving picture.} 
Large speed SO coupling band structures in the co-moving frame are plotted schematically for \textbf{a} $v_\mathrm{b}<0$ and \textbf{b} $v_\mathrm{b}>0$. 
Solid blue and red curves are bands for $\mid\uparrow\rangle$ and $\mid\downarrow\rangle$ without SO coupling while the dashed lines are SO-coupled bands with coupling strength $\Omega =1.53~E_\mathrm{r}$, the same as applied in the experiment. 
Initially, the ground state is in spin up component located in the left branch. 
Doppler shift in detuning is not plotted. 
SO coupling band structures in the co-moving frame for two different positive sweeping velocities \textbf{c} $v_\mathrm{b}=0.5$ mm s$^{-1}$ and \textbf{d} $v_\mathrm{b}=1.1$ mm s$^{-1}$. 
The dashed line denotes the energy from barrier kick. Only for $v_\mathrm{b}\gtrsim 1.1$ mm s$^{-1}$, can the atoms overcome the barrier in the lower band. }
\label{figs4}
\end{figure}

To understand the difference between these two critical velocities, we consider a co-moving frame with respect to the barrier, where the BEC moves at $-v_\mathrm{b}$ and is scattered by a static barrier. 
As a first approximation, consider a single particle scattered by a given barrier in the absence of SO coupling. 
Depending on the dynamical energy (set by the velocity) of the particle, it can either be transmitted or reflected. 
For the Gaussian barrier implemented in the experiment, the critical dynamical energy needed for the transition between reflection and transmission is $E_{\text{crit}}^{0}\sim 8.4E_\mathrm{r}$ (corresponding to a critical velocity $v_{\text{crit}}^{0}\sim 2.9v_\mathrm{r}\sim 16.8$ mm s$^{-1}$).

Now consider a moving barrier in the presence of SO coupling. 
The BEC is initially loaded into SO coupling with quasi-momentum $q_{i}$ and detuning $\delta$ in the lab frame. In the co-moving frame, this becomes $q_{i,\text{cm}}=q_{i}+\frac{v_\mathrm{b}}{v_\mathrm{r}}k_\mathrm{r}$ and $\delta _{\text{cm}}=\delta \pm \frac{4v_\mathrm{b}}{v_\mathrm{r}}E_\mathrm{r}$ for a left (+) or right (-) moving barrier, where the detuning change is due to the Doppler shift. 
The band structure in the co-moving frame is depicted in Supplemental Fig.~\ref{figs4}a,b. 
The dynamics here are mainly characterized by the lowest band in the presence of strong SO coupling ($\Omega =$1.53$E_\mathrm{r}$, as used in the experiment, is considered to be strong in this context). 
As a result, the effective dynamical energy for a left or right moving barrier is given by
\begin{equation}
E_\mathrm{D}^\mathrm{L}= \frac{(q_{i}+v_\mathrm{b}+1)^{2}}{v_\mathrm{r}^{2}}E_\mathrm{r}-2\frac{q_{i}}{k_\mathrm{r}}E_\mathrm{r}+\frac{\delta }{2}-E_\mathrm{r}\sqrt{(\frac{2q_{i}}{k_\mathrm{r}}-\frac{\delta }{2E_\mathrm{r}})^{2}+\frac{\Omega ^{2}}{4E_\mathrm{r}^{2}}}
\end{equation}
\begin{equation}
E_\mathrm{D}^\mathrm{R}= (\frac{4v_\mathrm{b}}{v_\mathrm{r}}E_\mathrm{r}-\delta )+\frac{(q_{i}-v_\mathrm{b}+1)^{2}}{v_\mathrm{r}^{2}}E_\mathrm{r}-2\frac{q_{i}}{k_\mathrm{r}}E_\mathrm{r}+\frac{\delta }{2}-E_\mathrm{r}\sqrt{(\frac{2q_{i}}{k_\mathrm{r}}-\frac{\delta }{2E_\mathrm{r}})^{2}+\frac{\Omega ^{2}}{4E_\mathrm{r}^{2}}}.
\end{equation}
The critical velocity for either direction is obtained by solving $E_\mathrm{D}^\mathrm{L,R}=E_{\text{crit}}^{0}$.
Based on the parameters used in the experiment, we find $v_{\text{crit}}^\mathrm{L}\simeq 16.6$~mm s$^{-1}$ and $v_{\text{crit}}^\mathrm{R}\simeq 9.3$~mm s$^{-1}$, which is consistent with numerical and experimental results. 

In the above discussion, we assume that at low speeds, the atoms are always able to overcome the momentum-space barrier in the lower SO coupling band such that the BEC is driven into the opposite spin state. 
In general, this is not true. 
Here, we define and determine a lower critical velocity limit for our system.
In Supplemental Fig.~\ref{figs4}c, the atoms are unable to overcome the momentum-space barrier due to small Doppler shift and small transferred energy, leaving the resonant coupling channel, $C_\mathrm{T_2}$, unattainable. 
As the barrier speed increases, the momentum-space barrier flattens, more energy is obtained from the barrier kick, and the resonant coupling channel opens. 
The lower critical velocity limit is found to be $\sim 1.1$ mm s$^{-1}$ (Supplemental Fig.~\ref{figs4}d) for experimental parameters, which is consistent with experimental and simulation results.

\subsection{Supplementary Note 4}
\label{sec:barrier-width-effects}

From the scattering dynamics, we determined that a large and wide barrier is desired for spin switching because a large parameter $\nu$ (Note \ref{sec:scatt-dynamics}) leads to a sharp transition between total reflection and total transmission and prevents an undesired mixture of reflection and transmission channels in a small range of velocities.
In addition, the barrier width may also affect how fast the atoms can be driven while still following the lower band, due to different Landau-Zener tunneling rates to a higher band for a given sweeping speed. 
In general, we expect to observe a zero tunneling rate in the `adiabatic' limit (i.e. with a very wide barrier). 
Consequently, the spin-flip transmission channel ($C_\mathrm{T_2}$) should be completely suppressed for a  sufficiently large barrier width, $w_\mathrm{b}$. 
To confirm this intuitive picture, we perform GPE simulations with different barrier widths for $v_\mathrm{b}=20$~mm s$^{-1}$. 
Results are shown in Supplemental Fig.~\ref{figs5}. 
We see that for $w_\mathrm{b}>7~\mu$m, $C_\mathrm{T_2}$ is completely suppressed. 
In the narrow barrier limit, where $C_\mathrm{T_2}$ is observable, atoms flipped into the $\mid \downarrow \rangle$ state (blue peaks in Supplemental Fig.~\ref{figs5} insets) have a negative final momentum. 
This behavior is in agreement with predictions from the resonant condition.

\begin{figure}[tbp]
\includegraphics[width=0.7\textwidth]{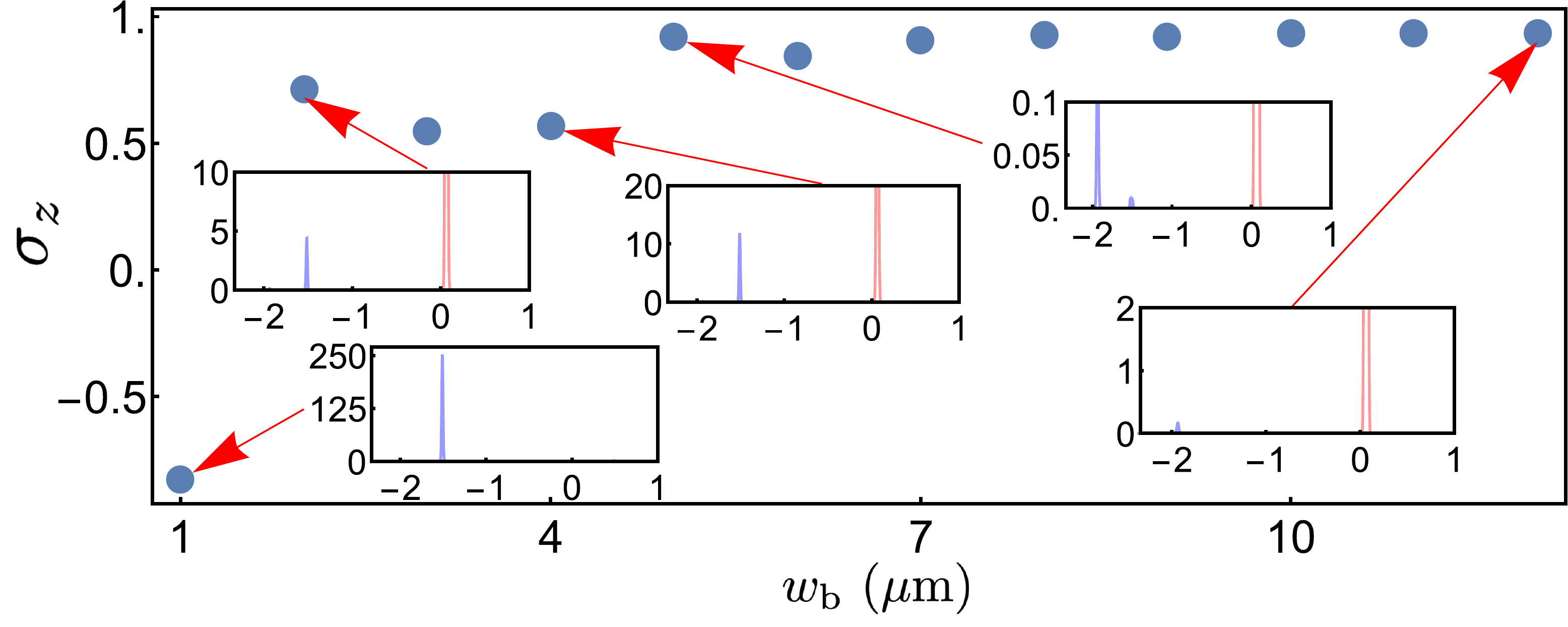}
\caption{\textbf{Barrier width dependence.} The spin polarization with respect to the barrier width in the single particle regime for a $v_\mathrm{b}=+20$~mm s$^{-1}$ sweep speed. Insets show corresponding momentum-space
profiles of final states. The spin flip process is accomplished through the $C_\mathrm{T_2}$ channel.}
\label{figs5}
\end{figure}

In conclusion, a reasonably wide barrier leads to a sharp transition in the transmission coefficient $T$ and with the help of avoided band crossing (i.e., strong SO coupling), suppresses the $C_\mathrm{T_2}$ channel. 
As a result, a good one-way spin switch should be expected under these parameters.



\begin{thebibliography}{50}
\bibitem{Ref0002} We note that small velocities correspond to long sweep
times and some of the observed changes in spin polarization at very low
velocities (i.e. $v_b < 2$~mm s$^{-1}$) may be attributed to heating in the system
induced in part by the Raman lasers.
\end{thebibliography}
\end{document}